%% file: main.tex
\documentclass[letterpaper,11pt,twoside]{article}

\usepackage{ifpdf}
\usepackage{microtype}
\usepackage{times}
\usepackage{graphicx}
\usepackage[table]{xcolor}
\usepackage{amsmath,amsfonts}
\usepackage[hyphens]{url}
\usepackage[numbers,sort]{natbib}
\usepackage{dcolumn,booktabs}
\usepackage{multirow}
\usepackage[compact]{titlesec}
\usepackage{fancyhdr}
\usepackage{shortcut}
\usepackage{wrapfig}
\usepackage{algorithm,algorithmic,rotating}
\usepackage{paralist}
\usepackage{xspace}
\usepackage{pifont}
\usepackage{url}
\usepackage{pgfgantt}
\usepackage{pgfplots}
\usepackage{tikz}
\usepackage{subfig}
\usepackage{hyperref}

 \hypersetup{
    colorlinks = true,
    linkcolor = red,
    anchorcolor = blue,
    citecolor = green,
    filecolor = blue,
    urlcolor = blue
}

\pdfoutput=1

\input{math_definition}

\usepackage{graphics}

\usepackage{color}

\ifpdf
  \pdfpagewidth=\paperwidth
  \pdfpageheight=\paperheight
\fi

\advance\textheight-\topmargin
\advance\textheight-\headheight
\advance\textheight-\headsep
\advance\textheight-\footskip
\advance\textheight12pt

\SetExpansion[context=sloppy,stretch=60,shrink=60,step=5]{encoding=*}{}

\newcolumntype{d}[1]{D{.}{.}{#1}}
\newcolumntype{R}[1]{D{/}{/}{#1}}

\usepackage{refcount}
 
\pgfplotsset{compat=1.13}
\begin{document}

\input{cover.tex}
\clearpage

\advance\baselineskip0pt plus.5pt minus0pt
\flushbottom

\pagenumbering{arabic}
\setcounter{page}{1}

\section{Introduction}
\label{s_intro}
\input{paper/1.introduction}

\section{Background: Compartmental Epidemic Models}
\label{s_bkg}
\input{paper/2.background}

\section{Methodology}
\label{s_approach}
\input{paper/3.approach}

\section{Evaluation of the Proposed Framework}
\label{s_exp}
\input{paper/4.exp}

\section{Related Work}
\label{s_related}
\input{paper/5.related}

\section{Conclusion}
\label{s_disc}
\input{paper/6.disc}

\bibliographystyle{ACM-Reference-Format}
\bibliography{main}

\end{document}

%% file: math_definition.tex
\newlength\savewidth\newcommand\shline{\noalign{\global\savewidth\arrayrulewidth\global\arrayrulewidth1pt}\hline\noalign{\global\arrayrulewidth\savewidth}}

\usepackage{amssymb}
\usepackage{amsmath,amsfonts}
\usepackage{xcolor}
\usepackage{amsopn}
\usepackage{bm} % bold symbol
\usepackage{multirow}
\usepackage{xspace}
\usepackage{subfig}

\newcommand{\ProbOpr}[1]{\mathbb{#1}}
\newcommand{\expect}[2]{%
\ifthenelse{\equal{#2}{}}{\ProbOpr{E}_{#1}}
{\ifthenelse{\equal{#1}{}}{\ProbOpr{E}\left[#2\right]}{\ProbOpr{E}_{#1}\left[#2\right]}}} % Expectation: syntax: E{1}{2} = E_1[2], E{}{2}=E[2], E{1}{} = E_1
\newcommand{\var}[2]{%
\ifthenelse{\equal{#2}{}}{\ProbOpr{VAR}_{#1}}
{\ifthenelse{\equal{#1}{}}{\ProbOpr{VAR}\left[#2\right]}{\ProbOpr{VAR}_{#1}\left[#2\right]}}} % Expectation: syntax: V{1}{2} = V_1[2], V{}{2}=V[2], V{1}{} = V_1
\newcommand{\eat}[1]{}

%% file: cover.tex
\thispagestyle{empty}

\noindent{}

\begin{center}

\vspace{0.4in}

\huge
\vspace{0.8cm}
        
{\LARGE \bf 
WLAN-Log-Based Superspreader Detection \\ in the COVID-19 Pandemic
}
\vfil

\normalsize
Cheng Zhang$^1$, Yunze Pan$^1$, Yunqi Zhang$^1$, Adam C. Champion$^1$,\\ Zhaohui Shen$^3$, Dong Xuan$^1$, Zhiqiang Lin$^1$, Ness B. Shroff$^{1,2}$\\

{$^{1}$Department of Computer Science and Engineering, 
 The Ohio State University\\
 $^{2}$Department of Electrical and Computer Engineering, 
 The Ohio State University\\
 $^{3}$VirtualKare LLC\\
}

\vfil
 
\end{center} 
\noindent 

\input{paper/0.abstract}

\par
 
\vfil

%% file: paper/0.abstract.tex
\begin{abstract}
Identifying ``superspreaders'' of disease is a pressing concern for society during pandemics such as COVID-19. Superspreaders represent a group of people who have much more social contacts than others. The widespread deployment of WLAN infrastructure enables non-invasive contact tracing via people's ubiquitous mobile devices. This technology offers promise for detecting superspreaders. In this paper, we propose a general framework for WLAN-log-based superspreader detection. In our framework, we first use WLAN logs to construct contact graphs by jointly considering human symmetric and asymmetric interactions. Next, we adopt three vertex centrality measurements over the contact graphs to generate three groups of superspreader candidates. Finally, we leverage SEIR simulation to determine groups of superspreaders among these candidates, who are the most critical individuals for the spread of disease based on the simulation results. We have implemented our framework and evaluate it over a WLAN dataset with 41 million log entries from a large-scale university. Our evaluation shows superspreaders exist on university campuses. They change over the first few weeks of a semester, but stabilize throughout the rest of the term. The data also demonstrate that both symmetric and asymmetric contact tracing can discover superspreaders, but the latter performs better with daily contact graphs. Further, the evaluation shows no consistent differences among three vertex centrality measures for long-term (i.e., weekly) contact graphs, which necessitates the inclusion of SEIR simulation in our framework. We believe our proposed framework and these results can provide timely guidance for public health administrators regarding effective testing, intervention, and vaccination policies. \\

\noindent\emph{Keywords}: Superspreader detection, WLAN logs, contact tracing, network analysis, COVID-19 pandemic

\end{abstract}

%% file: paper/1.introduction.tex
The COVID-19 pandemic has devastated many communities worldwide. The presence of the novel coronavirus (that causes COVID-19) in a community with high population density, such as a large public university, significantly increases the risk of contracting the disease. To fight COVID-19, contact tracing~\cite{eames2003contact,hellewell2020feasibility,klinkenberg2006effectiveness,salathe2020covid,singh2020blueprint} is especially important to discover active individuals, known as \emph{superspreaders},\footnote{There is no scientific definition of a ``superspreader''. We use a definition similar to that in \cite{reich2020modeling}: superspreaders are people with far more social connections than others, are more likely to be infected, and, if infected, will infect many more people than the median.} who lead to numerous COVID-19 transmission cases. Tracing human contacts to understand superspreader events is vital for preventing the spread of disease in communities such as university campuses, and such tracing has thus attracted a flurry of research interest~\cite{wen2020study,zhao2020accuracy,trivedi2020wifitrace,luo2020acoustic}.

Typically, contact tracing is conducted manually~\cite{cdc} (\eg, through questionnaires and
interviews), initially collecting necessary information from infected patients (such as locations they visited and people with whom they had contact). Unfortunately, manual contact tracing can result in inaccurate results due to people’s unreliable memories and long delays. Hence, to fight the COVID-19 pandemic, researchers have developed numerous (partially) automated contact tracing systems. Recent efforts can be divided into two categories: \emph{client-based} and \emph{infrastructure-based}. Client-based approaches require pervasive deployment of apps on people's mobile devices. Client-side apps leverage a wide variety of sources to track ``encounters,'' including records of credit card transactions~\cite{Pingbo}, cryptographic tokens exchanged via Bluetooth Low Energy (BLE)~\cite{covidwatch,pact,googleandapple,tracetogether}, or acoustic channels~\cite{luo2020acoustic}. In contrast, infrastructure-based methods exploit existing infrastructure deployed worldwide, such as CCTV footage~\cite{skoll2020covid}, locations measured using cellular networks~\cite{ayan2020characterizing}, Wi-Fi hotspots~\cite{trivedi2020wifitrace}, and GPS~\cite{bay2020bluetrace}, without requiring client-side involvement. \emph{In this context, our paper presents an approach leveraging Wi-Fi local area network (WLAN) logs to identify potential superspreaders on the campus of a large public university.}

However, leveraging WLAN logs for superspreader detection is nontrivial, with two major issues. {First}, conventional WLAN-based solutions (\eg, WiFiTrace~\cite{trivedi2020wifitrace}) infer whether students have contacted with each other based on their associations with specific access points (APs) during certain time intervals (\eg, $>15$ minutes). Such \emph{symmetric} contact detection neglects an important fact: the virus carried by people who have tested positive may infect others and replicate via pathogens in the environment. Therefore, others may be infected even if they linger in the environment over very short periods of time (\eg, $<15$ minutes). Obviously, the current definition of human contact cannot handle this scenario.  
{Second}, existing Wi-Fi-based methods~\cite{trivedi2020wifitrace} quantify a superspreader by the number of associated devices from the same access point. However, the number of contacts may be unable to truly reflect how critical an individual is to spreading disease amidst the population. For example, previous work on vertex centrality measurement for social network analysis~\cite{kolaczyk2014statistical} 
demonstrates that the ``importance'' of a specific vertex in message-passing not only depends on the number of connected vertices, but is relevant to the vertex's location in social networks. Moreover, ground truth remains unknown in WLAN-based contact graphs, making it hard to understand how fast the disease propagates and progresses in order to determine superspreaders.

To tackle the first issue, we introduce \emph{asymmetric contact}, a new type of human contact. Two persons in asymmetric contact are not necessarily associated with specific APs for the same period. For example, assume Persons A and B are in asymmetric contact. Person A may stay with one specific AP for a short time (e.g., $5$ minutes) whereas Person B stays longer ($50$ minutes). Due to Person B's longer stay time, he generates a much stronger ``environment'' with his microbes than Person A does. If B tests positive, he may infect A even if the latter's stay time is only $5$ minutes. On the other hand, A will not infect B due to her short stay. Hence the contact between these two persons is asymmetric. When we count the contact number, B's contact with A is counted, but A's contact with B is not. The concept of asymmetric contact partially captures the notion of \emph{environmental infection}\footnote{The physical environment represents an important source of pathogens that can cause infections or carry antibiotic resistance.}~\cite{environmental}. 

\begin{figure}[t]
    \centering
    \includegraphics[width=0.7\linewidth]{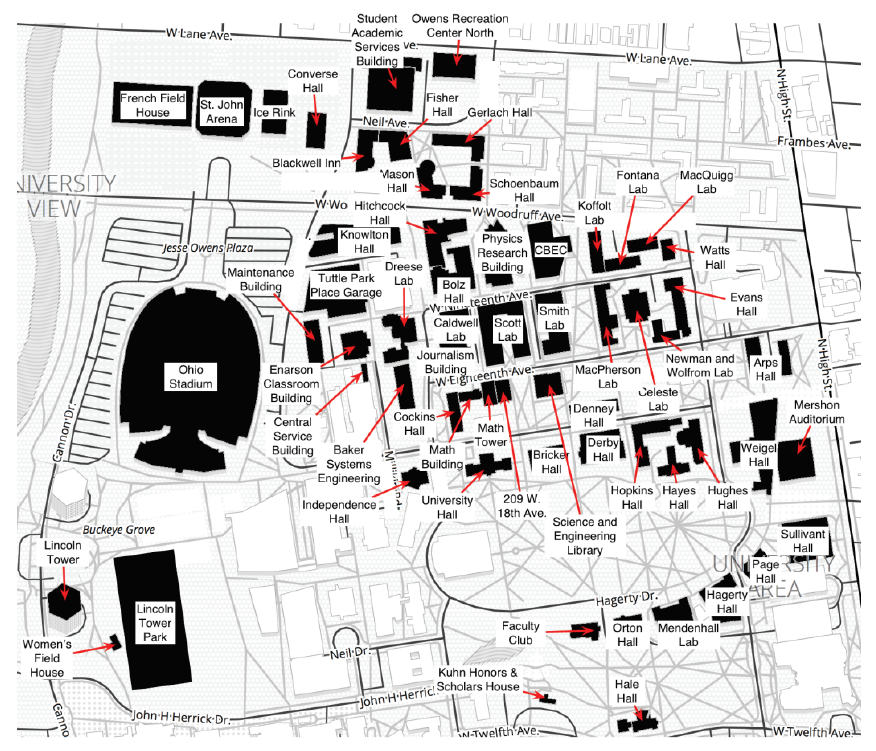}
    % \centerline{\fbox{\rule{0pt}{2in} \rule{0.9\linewidth}{0pt}}}
    \caption{\textbf{Campus buildings with AP deployment information (shaded)}. Other buildings include: 22 E. 16th Avenue, 53 W. 11th Avenue, Knight House, North Commons, Northwood-High Building, Raney Commons, Riverwatch Tower, and the Wexner Center for the Arts (not shown). We generate the map using Mapzen~\cite{mapzen} with OpenStreetMap data~\cite{openstreetmap}.}
    \label{fig:intro}
\end{figure}

As to the second issue, ideally, we can choose a vertex measure to determine superspreaders using either analytical solutions or prior experimental tests. Unfortunately, due to the diversity of contact graphs and the complexity of virus propagation, it is very difficult (if not impossible) to do so. In this paper, we propose an empirical approach. We include SEIR simulation, a necessary component in our solution, to ultimately determine superspreaders among the vertex-measure outputs. Specifically, we use the SEIR model to simulate the spread of the virus, followed by adaptive interventions on groups of superspreaders identified via different measures. We then finalize superspreaders who have the most crucial virus spread impacts, over the given contact graph, according to the simulations.

Incorporating the above two ideas, we propose a general framework for WLAN-log-based superspreader detection, which includes three key steps. 
{First}, we extract the individual's trajectory from wireless local area network (WLAN) logs to construct contact graphs, where vertices correspond to individual students and edges indicate physical contacts. In particular, we include both symmetric and asymmetric contact tracing to reveal potential directional interactions. 
{Second}, we adopt three vertex centrality measurements to identify three groups of potential superspreaders given a contact graph.
{Finally}, we leverage the SEIR model to compare different vertex centrality measures and determine superspreaders based on simulation results.

The WLAN dataset~\cite{cao2017human} used in this paper contains over 5,000 students at a large public university, which represents a random sample of the overall student body. Over 8,000 APs are deployed among more than 200 buildings on campus, including lecture halls, dormitories, and restaurants. There are over 41 million device (dis)associations with WLAN APs at the university over a 139-day observation period in 2015. Although insufficient WLAN logs are available in 2020 due to school closures originating from COVID-19, the 2015 logs describe campus interactions before. \autoref{fig:intro} shows the locations of APs in multiple buildings on campus.
Since the whole campus spans over 1,500 acres, our framework offers potential assisting superspreader detection efforts in large communities. 

% Although insufficient WLAN logs are available in the year 2020 due to school closures by COVID-19, the patterns of students’ mobility in 2015 share similarities to those in the years afterward. \autoref{fig:intro} shows the locations of APs in multiple buildings on campus. Since the whole campus spans over 1,500 acres, our findings offers potential assisting superspreader detection efforts in large communities.

% \Cheng{yes, we need to edit this paragraph}
% \Adam{Yes, agreed!}
% Although insufficient WLAN logs are available in 2020 due to school closures originating from COVID-19, the 2015 logs describe campus interactions before . 
% Since the whole campus spans over 1,500 acres, our framework  offers potential assisting superspreader detection efforts in large communities.

{\textbf{The main findings of this work include the following:}} (1) We find that there is a group of students that is critical in spreading the virus throughout the university's social contact networks. (2) We show the importance of symmetric and asymmetric contact tracing in superspreader detection. Specifically, we show that asymmetric contact tracing helps to discover hidden superspreaders in daily contact graphs and proper interventions with identified superspreaders greatly boosts efforts to contain the spread of disease. (3) We find that simple \emph{betweenness} centrality better reveals the most critical individuals in daily contact graphs. We do not observe notable differences between vertex centrality measures in longer-term (\emph{i.e.,} weekly) contact graphs with epidemic control. (4) For resource-constrained quarantine, we observe that increasing the percentage of the quarantined individuals to over 20\% of the population yields limited extra benefits. (5) We find that superspreaders change heavily over the first few weeks, then remain stable during the rest of the semester. The similarity of superspreaders between the first 20 weeks and 15 weeks is around 0.8 using the rank-biased overlap metric~\cite{webber2010similarity}, opening up opportunities to discover superspreaders as early as possible for efficient pandemic mitigation.

{\textbf{Practical significance for university/city administrators:}} 
We believe our proposed contact tracing method will enable both proactive and reactive interventions. For the former, our method can help administrators rapidly identify superspreaders for health warnings and frequent testing, using data from just the first few weeks of the semester. For the latter, our method can assist efforts in contact tracing, quarantine, medical support, and prioritized patient care.

In summary, our main contributions are threefold:
\begin{itemize}
    \item We propose a general framework for WLAN-log-based contact analysis and superspreader detection. The framework applies to a wide range of working scenarios based on users' preferences, environmental dynamics, and resource availability.
 
    \item We present a set of initial work using the WLAN-log-based superspreader detection framework, including asymmetric contact tracing, vertex centrality measurement, and simulation-based superspreader determination. 
    
    \item We implement the framework and evaluate it on a large-scale real-world WLAN log dataset. Our empirical results show the efficacy of the proposed contact tracing approaches and uncover insightful findings for public health administrators.
\end{itemize}

The rest of this paper is organized as follows. 
Section~\ref{s_bkg} provides background on epidemic models. 
%Section.~\ref{s_overview} gives an overview of the proposed framework. 
Section~\ref{s_approach} presents our framework on WLAN-log-based superspreader detection. 
Section~\ref{s_exp} illustrates our evaluation results and analyses. 
Section~\ref{s_related} reviews related work. 
Finally, Section~\ref{s_disc} concludes the paper.

%% file: paper/2.background.tex
In this section, we discuss the background of compartmental epidemic models, which are simplified mathematical models of infectious diseases~\cite{kermack1927contribution,hethcote1989three,anderson1992infectious,hethcote2000mathdisease,vynnycky2010introduction}. Recently, the SEIR (\textbf{S}usceptible, \textbf{E}xposed, \textbf{I}nfectious, \textbf{R}ecovered) model has shown promise combating COVID-19 in disease modeling~\cite{ferguson2020impact,reich2020modeling}, forecasting~\cite{bertozzi2020challenges,van2020forecasting}, and intervention~\cite{leung2020contact}. In the SEIR model, the population is assigned to labeled compartments between which people move based on their health status. 

\begin{figure}[t]
    \centering
    \includegraphics[width=0.7\linewidth]{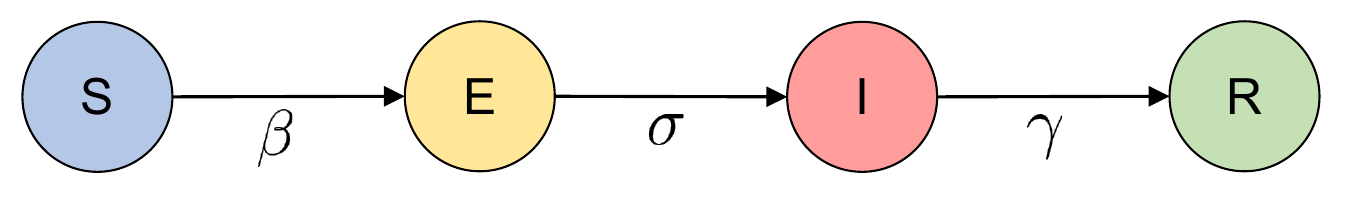}
    % \centerline{\fbox{\rule{0pt}{2in} \rule{0.9\linewidth}{0pt}}}
    \caption{\textbf{Illustration of the popular SEIR compartmental model in epidemiology.} The population is assigned to one of several labeled compartments: Susceptible, Exposed, Infectious, or Recovered. The order of the labels usually shows flow patterns between the compartments with epidemiological parameters $\beta$, $\sigma$, and $\gamma$. Details are explained in the text.}
    \label{fig:seir}
\end{figure}

Following the equivalent compartmental diagram shown in \autoref{fig:seir}, we can use the following differential equations to describe the SEIR model involving variables $S$, $E$, $I$, and $R$ and their rates of change with respect to time $t$:
\begin{equation}
    \frac{dS}{dt}=-\beta\frac{IS}{N}, \qquad
    \frac{dE}{dt}=\beta\frac{IS}{N}-\sigma E, \qquad
    \frac{dI}{dt}=\sigma E-\gamma I,  \qquad  
    \frac{dR}{dt}=\gamma I,\\
\end{equation}\label{eq:seir}
\noindent where $\beta$ is the probability of transmitting disease between a susceptible and an infectious individual, $\sigma$ is the inverse of the average incubation time (the rate of latent individuals becoming infectious), and $\gamma$ is the recovery rate.  In this model, recovered individuals are permanently immune to disease. In practice, all parameters are constant values that can be obtained via maximum likelihood estimation with real pandemic data. In this work, we leverage SEIR simulations~\cite{reich2020modeling} to model quarantine (self-isolation) of identified superspreaders (cf. \autoref{ss_setup}).

%% file: paper/3.approach.tex
\begin{figure}[t]
    \centering
    \includegraphics[width=1\linewidth]{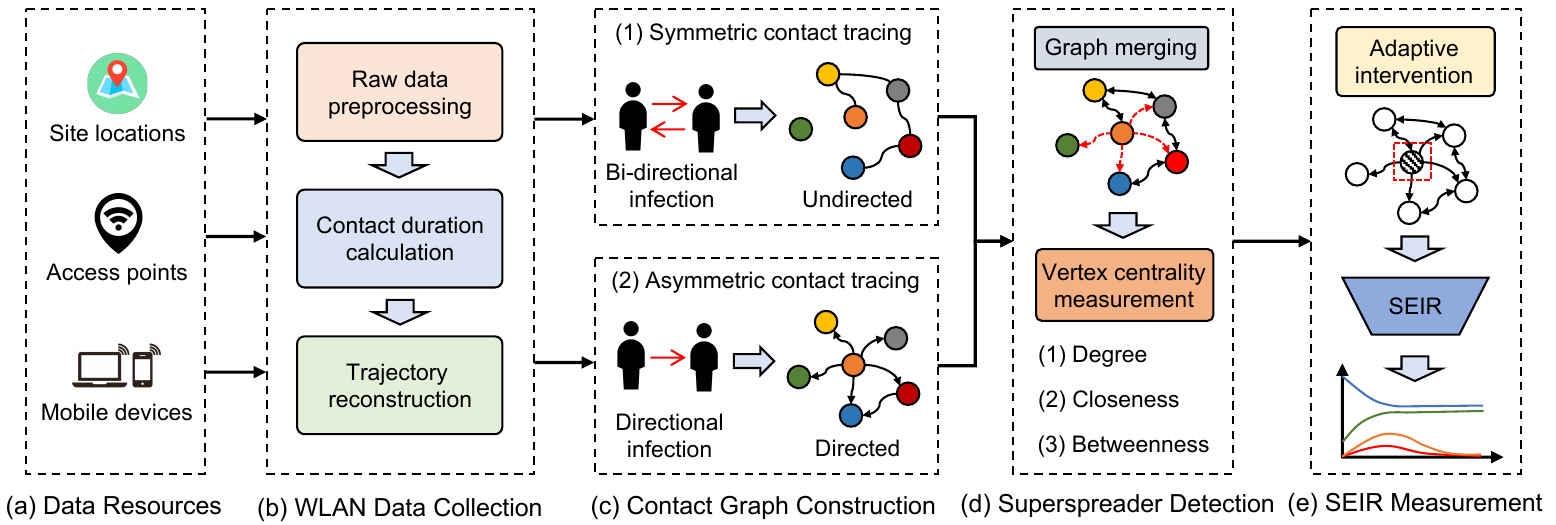}
    % \centerline{\fbox{\rule{0pt}{2in} \rule{0.9\linewidth}{0pt}}}
    \caption{\textbf{Overview of WLAN-log-based superspreader detection.} First, we extract contact graphs from WLAN logs via symmetric and asymmetric contact tracing. Second, we perform vertex centrality measurement to discover potential superspreaders. Finally, we simulate adaptive interventions using the SEIR model.}
    \label{fig:framework}
\end{figure}

\autoref{fig:framework} shows an overall framework of our WLAN-log-based superspreader detection. We describe each component as follows.

\subsection{WLAN Data Collection}
The WLAN logs often include the (dis)association of mobile devices with respect to APs. In this paper, we use the same dataset in \cite{cao2017human}. A sample log entry has the following format:

\smallskip
\noindent{\texttt{timestamp,process,ap-name,student-id,role,MAC,SSID,result}}
\smallskip

The fields in the log represent the event’s UNIX timestamp, the process that generated the log entry, the AP name, the encrypted student ID, the role assigned to the device, the anonymized MAC address (preserving the OUI), the
SSID name, and the authentication result (success or failure), respectively. 

Our WLAN dataset collection consists of three steps: (1) We first filter out students who use the university's unsecured WLAN from the dataset. Some information is missing regarding student ID and AP's name. We consider these log entries invalid in this work. After removing invalid entries from the dataset, 39 million log entries remain. (2) Since WLAN logs only provide the association (arrival) time of the person at the corresponding AP, we need to estimate the disassociation (leave) time. We first sort the log entries of each student in ascending order (based on timestamps) to ensure sequential order. For APs within the same building, the stay time of each AP is the duration between the arrival time of the next AP and the current one. Following \cite{cao2017human}, we also calculate the estimated walking time between two buildings using the Google Maps API~\cite{googlemap}. (3) In~\cite{cao2017human}, the location granularity is building-level as that work focuses on human mobility measurement \cite{kim2006extracting}. In contrast, we treat the AP as the base unit in the trajectory in order to study human proximity tracing. Therefore, after data processing, each user/MAC’s trajectory becomes a time series of APs and their corresponding stay times. A person's trajectory $T$ can be defined as:
\begin{equation}
    T = (AP_1, t_1, ST_1) \rightarrow (AP_2, t_2, ST_2) \rightarrow \dots \rightarrow (AP_M, t_M, ST_M), \\ t_1 < t_2 < \dots < t_M, \notag
\end{equation}
where $AP_i$ is the $i$th AP in trajectory $T$, $t_i$ is the arrival time of the person at $AP_i$, and $ST_i$ is the stay time of the person at $AP_i$. We refer to $(AP_i, t_i, ST_i)$ as a \emph{tracklet}. \autoref{fig:time} (top) shows how we estimate stay time for intra- and inter-building AP connections for a person's trajectory and illustrate Bob's trajectory between two buildings. 

\subsection{Contact Graph Construction}
\begin{figure}[t]
    \centering
    \includegraphics[width=0.95\linewidth]{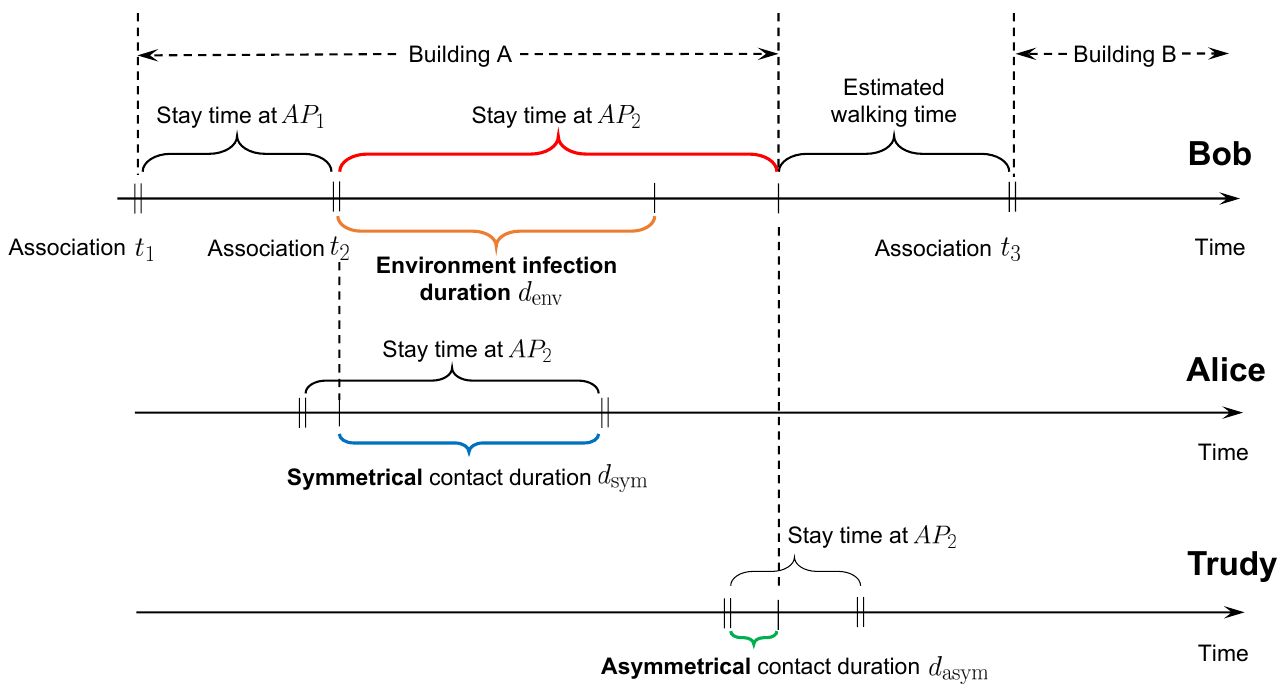}
    % \centerline{\fbox{\rule{0pt}{2in} \rule{0.9\linewidth}{0pt}}}
    \caption{\textbf{Contact tracing using persons' trajectories.} We show trajectories of three persons, \ie, Bob, Alice, and Trudy. At $AP_2$, Bob's stay time {\color{red}(red)} is longer than the environmental infection duration {\color{orange}(orange)}. There is a symmetric contact {\color{blue}(blue)} between Bob and Alice and an asymmetric contact {\color{green}(green)} between Bob and Trudy.} 
    \label{fig:time}
\end{figure}

Next, we describe the contact tracing method using persons' trajectories. Given a student's trajectory $T$ with sequential tracklets, we take each tracklet as a query and apply beam search on all other students' tracklets to determine if there is an overlapping duration for physical interaction between two persons. \autoref{fig:time} shows an example where we consider two contact tracing methods---symmetric and asymmetric---to compute the overlapping duration between Bob and two other persons, Alice and Trudy. 

\paragraph{{Symmetric contact tracing}}
Intuitively, if Bob and Alice connect to the same AP with a certain overlapping period, we assume there may be a potential physical interaction between them. Thus, given a tracklet $(AP_q, t_q, ST_q)$ from student $q$ (\eg, Bob) and a tracklet $(AP_p, t_p, ST_p)$ from student $p$ (\eg, Alice), we assign a bidirectional\footnote{The virus can spread from person $q$ to person $p$, and vice versa.} contact edge between $q$ and $p$ if $AP_q = AP_p$ and the following criterion is satisfied:
\begin{equation}
ST_q + ST_p - \max \{t_q + ST_q, t_p + ST_p\} + \min \{t_q, t_p\} \geq d_\text{sym},
\end{equation}
where $d_\text{sym}$ is a constant value of symmetric contact duration. Empirically, we set $d_\text{sym}$ to 15 minutes in the experiments.

\begin{figure}[t]
    \centering
    \includegraphics[width=0.8\linewidth]{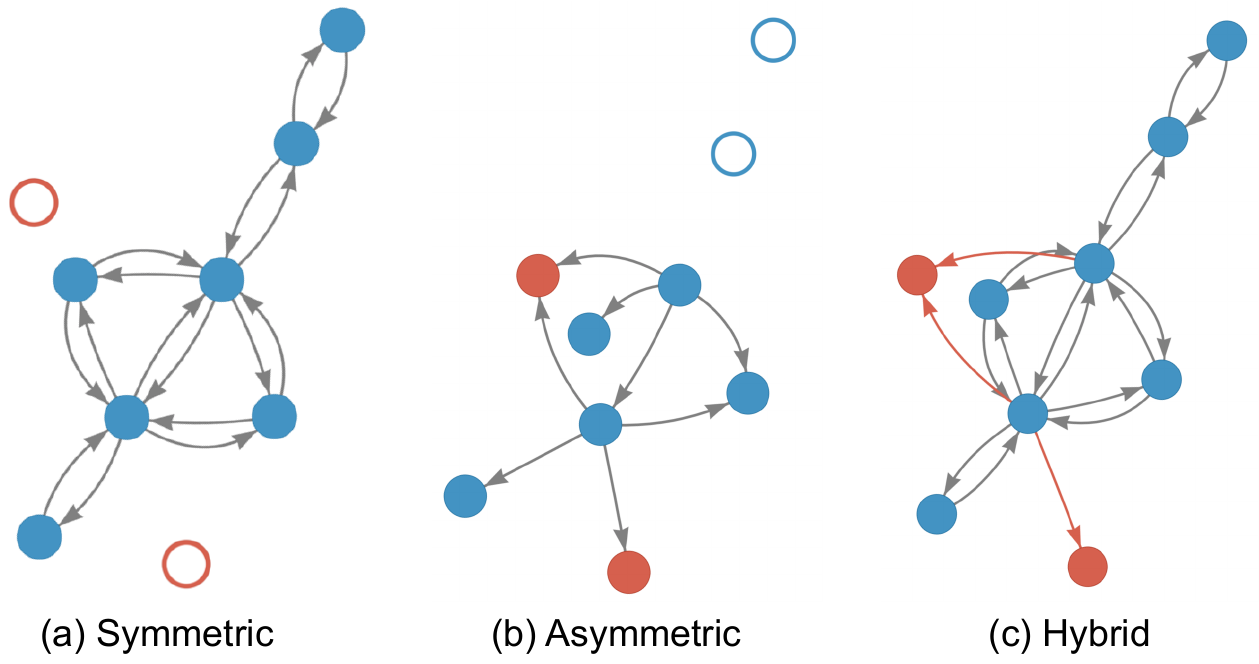}
    % \centerline{\fbox{\rule{0pt}{2in} \rule{0.9\linewidth}{0pt}}}
    \caption{\textbf{Symmetric, asymmetric, and hybrid contact graphs.} We show different contact tracing results of a real case from a group of students in our WLAN dataset. (a) Contact graph only with symmetric tracing: the \emph{unfilled} {red} nodes are overlooked due to short overlapped stay time with other blue nodes. (b) Contact graph with asymmetric tracing: we observe that \emph{filled} {red} nodes are included if directional contact is considered. (c) Merging symmetric and asymmetric graphs to construct a hybrid graph: red nodes and edges indicate newly discovered information compared to the symmetric contact graph.}
    \label{fig:contact_tracing}
\end{figure}

\paragraph{{Asymmetric contact tracing}}
However, the above symmetric tracing method omits environmental infection (cf. \autoref{s_intro}). In that situation, Bob may stay at $AP_2$ for a long enough period, making the environment infected. Thus, the virus will spread to another person, Trudy, even though the overlapping contact duration is short. To resolve this problem, we propose a new asymmetric contact tracing method that can discover such directional interactions. Concretely, we take Bob's tracklet whose stay time $ST_q$ exceeds a certain duration $d_\text{env}$ and assign a directional\footnote{The virus may only spread from one person to another.} contact edge between $q$ (\eg, Bob) and $p$ (\eg, Trudy) if $AP_q = AP_p$ and the following criterion is satisfied:
\begin{equation}
(ST_q - {\color{red}d_\text{env}}) + ST_p - \max \{t_q + ST_q, t_p + ST_p\} + \min \{t_q + {\color{red}d_\text{env}}, t_p\} \geq d_\text{asym} ,
\end{equation}
where $d_\text{env}$ and $d_\text{asym}$ are constant values of environmental infection time and asymmetric contact duration, respectively. Empirically, we set $d_\text{env}$ to 50 minutes and $d_\text{asym}$ to 5 minutes in the experiments. 

\paragraph{Graph Merging}
Once both symmetric and asymmetric contact graphs are obtained, we merge two graphs into one hybrid graph by aligning nodes and edges. The hybrid graph can reveal realistic contacts in our social interactions evidenced by WLAN logs. \autoref{fig:contact_tracing} gives an example for each graph. 

\subsection{Superspreader Detection via Vertex Centrality Measurement}
\label{ss_centrality}
\begin{figure}[hp]
    \centering
    \includegraphics[width=0.9\linewidth]{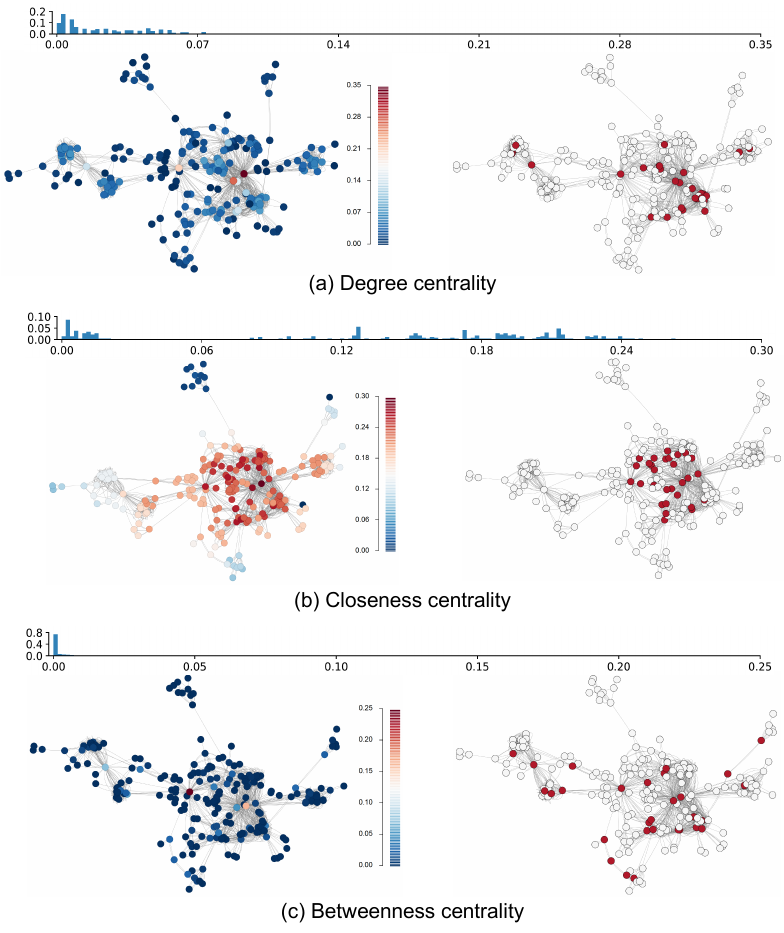}
    % \centerline{\fbox{\rule{0pt}{2in} \rule{0.9\linewidth}{0pt}}}
    \caption{\textbf{Visualization of vertex centrality measurement}. We show a one-day contact graph of a building on campus with (a) degree centrality, (b) closeness centrality, and (c) betweenness centrality measurements. The top, left, and right part of each indicates the relative frequency histogram, centrality graphs, and the top 10\% of highlighted nodes ({\color{red}{red}}), respectively. Warmer colors indicate larger values. Discrepancies among the three measurements are visible.}
    \label{fig:centrality}
\end{figure}

The reader may ask a key question about a vertex in the hybrid graph: \emph{How ``important'' is a specific person in the spread of disease?} Centrality measurements~\cite{kolaczyk2014statistical} are designed to quantify a person's importance, helping answer this question. Accordingly, the purpose of this subsection is \emph{not} to propose a new metric for vertex measurement. Rather, we investigate the efficacy of three metrics in representing superspreaders in the Wi-Fi-based contact graphs. \autoref{fig:centrality} shows their differences.

\paragraph{Degree centrality}
Degree centrality is defined as the number of edges incident upon a vertex (\ie, the vertex's number of social ties). If the network is directed, then two separate measures of degree centrality are defined: in-degree and out-degree. In this paper, we define each vertex's out-degree as follows:
\begin{equation}
    deg(u) = \frac{|E^{o}_{u}|}{N-1},
\end{equation}
where $|E^{o}_{u}|$ is the total number of edges directed out of a vertex $u$ in a directed hybrid contact graph, and $N$ is the number of vertices in the graph.

\paragraph{Closeness Centrality}
One common notion of centrality is a vertex's ``nearness'' to many other vertices, which closeness centrality metrics aim to capture. For a given vertex, closeness centrality varies inversely with the vertex's distance of a vertex from all others. Formally, for a connected graph, this measure is defined as:
\begin{equation}
    cl(u) = \frac{1}{\sum_{v} dist(u, v)},
\end{equation}
where $dist(u, v)$ denotes the geodesic (shortest-path) distance between vertices $u$ and $v$. Intuitively, this measure looks at how fast information can spread from one vertex to all others. For example, a vertex that is close to many other vertices may easily transmit the disease to them.

\paragraph{Betweenness centrality}
Another popular class of centralities is based upon the perspective that ``importance'' relates to a vertex's position regarding paths in the graph. If we picture those paths as the routes by which communication takes place, vertices situated on many paths tend to be more critical to the communication process. Betweenness centrality metrics are aimed at summarizing the extent to which a vertex is located ``between'' other pairs of vertices:
\begin{equation}
    bw(u) = \sum_{s \not= t \not= v} \frac{\sigma(s,t|v)}{\sigma(s, t)},
\end{equation}
where $\sigma(s,t|v)$ is the total number of shortest paths between $s$ and $t$ that pass through $v$ and $\sigma(s,t)=\sum_{v}\sigma(s,t|v)$. Vertices with high betweenness centrality are critical for maintaining graph  connectivity.

\paragraph{SEIR Measurement}
Based on these centrality measures, we are able to identify potential superspreaders. Next, we perform adaptive interventions on those active nodes using SEIR simulations to measure who are the most critical individuals for the spread of disease based on the simulation results.

%% file: paper/4.exp.tex
% !TEX root=main.tex

In this section, we first describe our methodology. Next, we present our experimental results.

\subsection{Evaluation Setup}
\label{ss_setup}
\paragraph{WLAN dataset}
We use the WLAN dataset from Cao et al.~\cite{cao2017human}, which contains WLAN log data with demographic information at a large public university spanning 139 days in 2015. Cao et al.~\cite{cao2017human} found that university students' mobility patterns change periodically on a weekly basis. In our study, we focus on analyzing the contact graph from a specific day of the week in the dataset. Specifically, we use the WLAN log information to compute the contact graph for each weekday from a randomly selected week in the dataset. We also report results on the contact graphs computed from a weekly period. 
We construct three types of contact graphs: symmetric, asymmetric, and hybrid.

\paragraph{Evaluation metrics}
Based on the SEIR model, we use the following realistic epidemiological measures to estimate the effect of different approaches:
\begin{itemize}
    \item \textbf{Doubling Time (day):} the time it takes for the number of cumulative infections to double.
    \item \textbf{Total Infected Fraction (\%)}: the fraction of the total accumulated infected population during the entire epidemic.
    \item \textbf{Peak Infected Time (day)}: the time required to infect the largest possible population.
    \item \textbf{Peak Infected Fraction (\%)}: the fraction of infected persons when peak infection is reached.
\end{itemize}

\paragraph{Experimental comparison}
We quarantine persons with higher centrality based on the hybrid contact graph and simulate the epidemic on the hybrid graph. We test three vertex centrality measurement methods and compare our results to the following baselines:

\begin{itemize}
    \item \textbf{No quarantine}: we let the virus spread naturally on the hybrid graph without intervention. 
    \item \textbf{Random}: we randomly quarantine a certain number of persons and simulate the epidemic on the hybrid graph.  
    \item \textbf{Symmetric contact tracing (SymC)}: we quarantine persons with higher centrality based on the symmetric contact graph and simulate the epidemic on the hybrid graph.
    \item \textbf{Symmetric and asymmetric contact tracing (Hybrid)}: we quarantine persons with higher centrality based on the hybrid contact graph and simulate the epidemic on the hybrid graph.
\end{itemize}

\paragraph{Implementation details}
We follow \cite{reich2020modeling} in order to simulate an epidemic using the SEIR model. We use the default SEIR parameters, as they are calculated from a real-world infectious dataset.\footnote{Other toolkits could be used to simulate the spread of disease elsewhere.} In particular, the total population size in the our experiments is 3748. We set the initial number of infected persons to 50, which we fix across all experiments. In order to achieve stable observations, We run our simulation 50 times in each group of experiments until convergence is reached.

\begin{table}[]
\centering
\caption{\textbf{Main results on single-day contact graph.} We compare different methods with various centrality metrics. Next, we perform SEIR simulation by quarantining 100 persons based on these metrics. We observe that our hybrid graph, which jointly considers symmetric and asymmetric contact tracing, achieves better performance than the baseline model and the symmetric contact tracing method alone. \textbf{DB-Time:} Doubling Time (day); \textbf{T-Inf:} Total Infected Fraction (\%); \textbf{PK-Time:} Peak Infection Time (day); \textbf{PK-Inf:} Peak Infection Fraction (\%). Results in {\color{blue}blue} show where the hybrid graph outperforms SymC. The top result in each column is in \textbf{bold}.}
\begin{tabular}{lcrrrr}
\shline
\multicolumn{1}{c}{Method} & Measure & \multicolumn{1}{c}{DB-Time ($\uparrow$)} & \multicolumn{1}{c}{T-Inf ($\downarrow$)} & \multicolumn{1}{c}{PK-Time ($\uparrow$)} & \multicolumn{1}{c}{PK-Inf ($\downarrow$)} \\
\hline
No quarantine & - & 3.24 & 48.45 & 29.00 & 4.17 \\
Random & - & 3.29 & 44.90 & 30.40 & 3.91 \\
\hline
\hline
SymC & \multirow{2}{*}{Degree} & 5.61 & 40.69 & 40.80 & 2.53 \\
Hybrid  &  & {(-0.70)} 4.91 & {\color{blue}(-1.92)} 38.77 & {(-0.60)} 40.20 & {\color{blue}(-0.27)} \textbf{2.26} \\
\hline
SymC & \multirow{2}{*}{Closeness} & 6.08 & 40.21 & 39.80 & 2.37 \\
Hybrid &  & {\color{blue}(+0.51)} \textbf{6.59} & {\color{blue}(-1.74)} 38.47 & {\color{blue}(+1.20)} 41.00 & {\color{blue}(-0.09)} 2.28 \\
\hline
SymC & \multirow{2}{*}{Betweenness} & 5.44 & 42.61 & 39.80 & 2.51 \\
Hybrid &  & {\color{blue}(+0.21)} 5.65 & {\color{blue}(-5.20)} \textbf{37.41} & {\color{blue}(+1.40)} \textbf{41.20} & {\color{blue}(-0.06)} 2.45 \\
\shline
\end{tabular}
\label{tab:main_results}
\end{table}

\begin{figure}[t]
    \centering
    \includegraphics[width=0.5\linewidth]{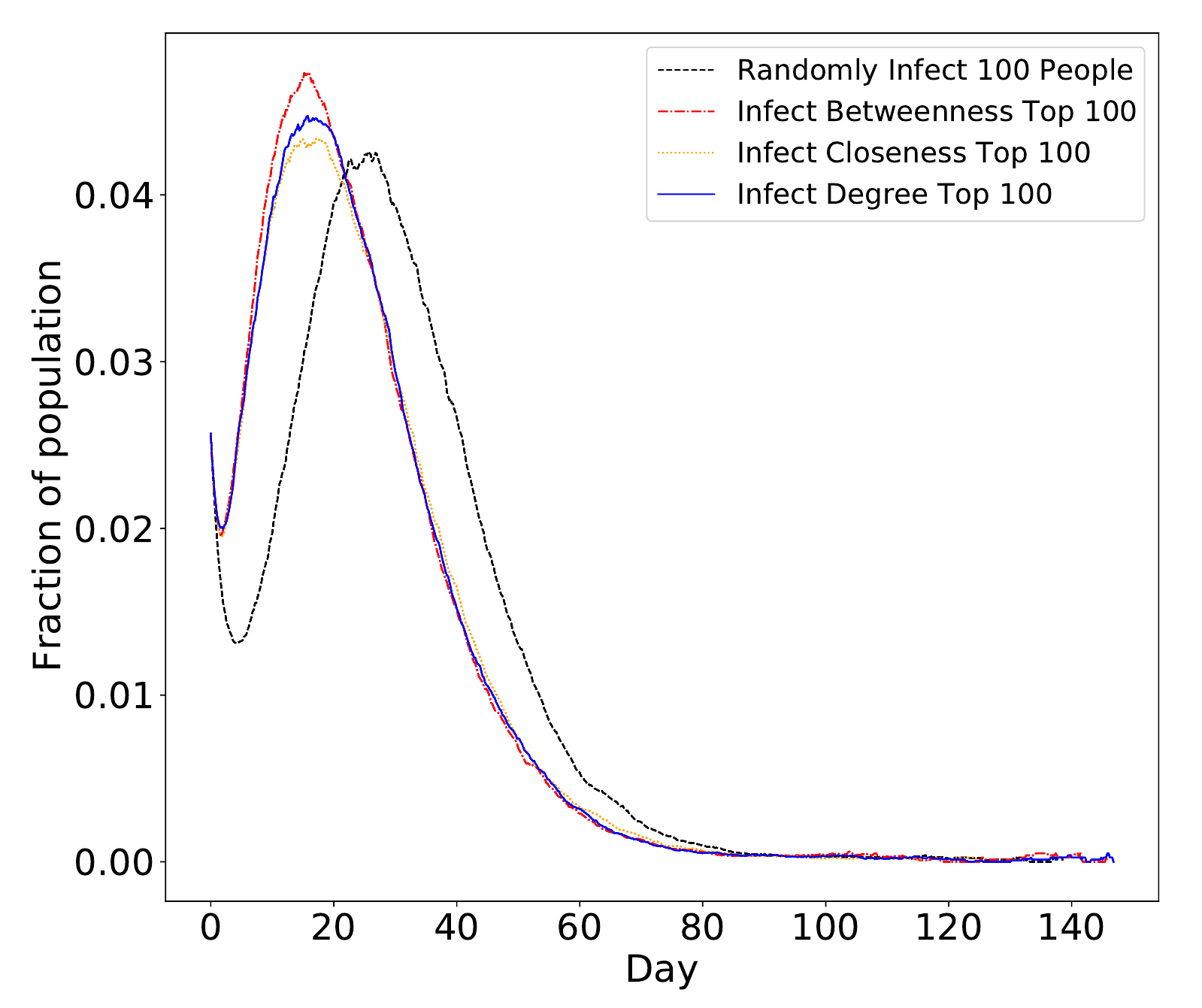}
    % \centerline{\fbox{\rule{0pt}{2in} \rule{0.9\linewidth}{0pt}}}
    \caption{\textbf{Effect of the infected population on the spread of the pandemic.} We select 100 students and set their initial conditions as infectious based on different criteria. We run SEIR simulation and show the fraction of the infected population on different days.}
    \label{fig:exist}
\end{figure}

\begin{figure}[th]
    \centering
    \includegraphics[width=0.98\linewidth]{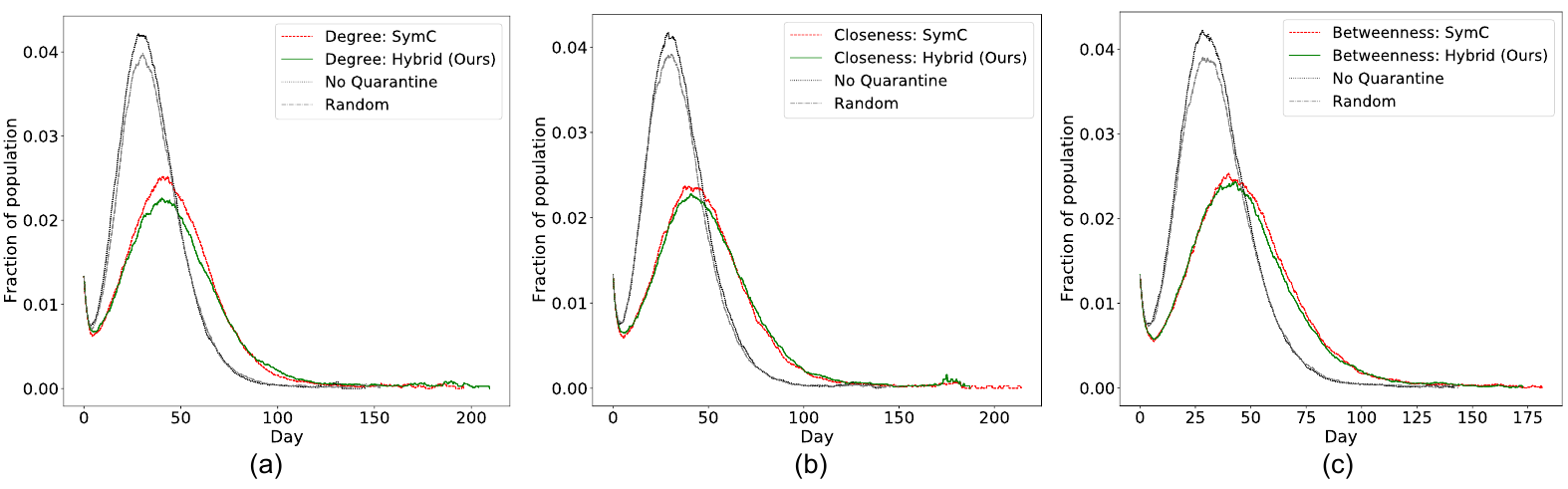}
    % \centerline{\fbox{\rule{0pt}{2in} \rule{0.9\linewidth}{0pt}}}
    \caption{\textbf{Spread of the pandemic during the period.} We show comparison results on (a) degree centrality, (b) closeness centrality, and (c) betweenness centrality measurements. Our asymmetric contact tracing and symmetric contact tracing ({{\color{green}{green}} and \color{red}{red}}) outperforms the baseline approaches with random quarantine {\color{gray}{(gray)}}.}
    \label{fig:curve}
\end{figure}

\subsection{Results and Analyses}
\paragraph{Main results on single-day contact graph}
We report the main results of a single-day contact graph in \autoref{tab:main_results}. Identifying superspreaders using a hybrid graph with asymmetric and symmetric contact tracing outperforms baseline methods substantially in terms of all centralities, justifying our motivation: \emph{symmetric and asymmetric contact tracing, which naturally reflects environmental infection, can be a valuable factor to contain the spread of disease.} In addition, we find similar observations from other days of the week in the WLAN dataset. Next, we detail our analyses.\smallskip

\noindent\textbf{Superspreaders exist on the university campus.}
We notice that both SymC and Hybrid significantly outperform baseline and ``random quarantine,'' suggesting the existence of superspreaders and the importance of contact tracing to limit the spread of disease. To analyze these superspreaders' extent of spread, we conduct a simulated comparison by initializing different groups of individuals. As shown in \autoref{fig:exist}, we observe that the virus carried by students with higher centrality causes a much faster spread than with randomly selected students. Further, students with higher betweenness centrality are critical to the spread.\smallskip

\noindent\textbf{Asymmetric contact tracing is efficient.}
We found that asymmetric contact tracing with a simple vertex measure leads to a notable gain for all metrics. Especially for the total infected fraction (T-Inf), Hybrid is $\sim$1\% better than symmetric contact tracing (SymC), which represents around 40 persons in our WLAN dataset. We also show the SEIR simulation curves in \autoref{fig:curve}: both symmetric and asymmetric contact tracing methods significantly outperform random quarantine methods, demonstrating the effectiveness of our superspreader detection framework.\smallskip

\noindent\textbf{Betweenness centrality strongly limits the total infected population on daily contact graphs.}
By comparing different centrality measurements for the selection of quarantine populations, we found that betweenness centrality leads to the strongest reduction in the total infected fraction (from $42.61\%$ to $37.41\%$) in the daily contact graph (cf. \autoref{tab:main_results}). One reason is that betweenness metrics can effectively discover vertices that sit on many paths are likely more critical to the spread process in social graphs. This verifies our observation in \autoref{fig:centrality} that betweenness centrality identifies a very different group of persons compared to degree centrality and closeness centrality (cf. \autoref{ss_centrality}).

\begin{table}[t]
\tabcolsep 10pt
\centering
\caption{\textbf{Results on one-week contact graph.} We compare different methods using a one-week contact graph. We perform SEIR simulation by quarantining 100 persons based on centrality metrics. We use betweenness centrality to discover superspreaders. \textbf{CM:} Centrality Measure; \textbf{DB-Time:} Doubling Time (day); \textbf{T-Inf:} Total Infected Fraction (\%); \textbf{PK-Time:} Peak Infection Time (day); \textbf{PK-Inf:} Peak Infection Fraction (\%).}
% \resizebox{\columnwidth}{!}{%
\begin{tabular}{lccccc}
\shline
\multicolumn{1}{c}{Method} & \multicolumn{1}{c}{DB-Time ($\uparrow$)} & \multicolumn{1}{c}{T-Inf ($\downarrow$)} & \multicolumn{1}{c}{PK-Time ($\uparrow$)} & \multicolumn{1}{c}{PK-Inf ($\downarrow$)} \\
\hline
No quarantine  & 0.98 & 86.15 & 13.04 & 17.17 \\
Random  & 0.98 & 83.93 & 13.04 & 16.61 \\
\hline
\hline
SymC & 1.09 & 81.88 & {13.96} & {16.18} \\
Hybrid & {1.11} &  {82.57} &  13.84 & 16.17 \\
\shline
\end{tabular}%
\label{tab:week}
% }
\end{table}

Further, we extend the simulation on contact graphs computed over longer weekly periods. Compared to daily contact graphs, weekly graphs generated from the WLAN logs are more densely connected. We focus on betweenness centrality and \autoref{tab:week} shows the results. We find that the difference between the symmetric and the hybrid graphs is marginal. This is because, in long-term contact tracing, the top superspreaders between the symmetric and hybrid graphs overlap highly, suggesting that early-stage pandemic control helps identify superspreaders who may be missed otherwise. We observe similar patterns for other weeks throughout the study period.

\begin{figure}[h]
    \centering
    \includegraphics[width=0.55\linewidth]{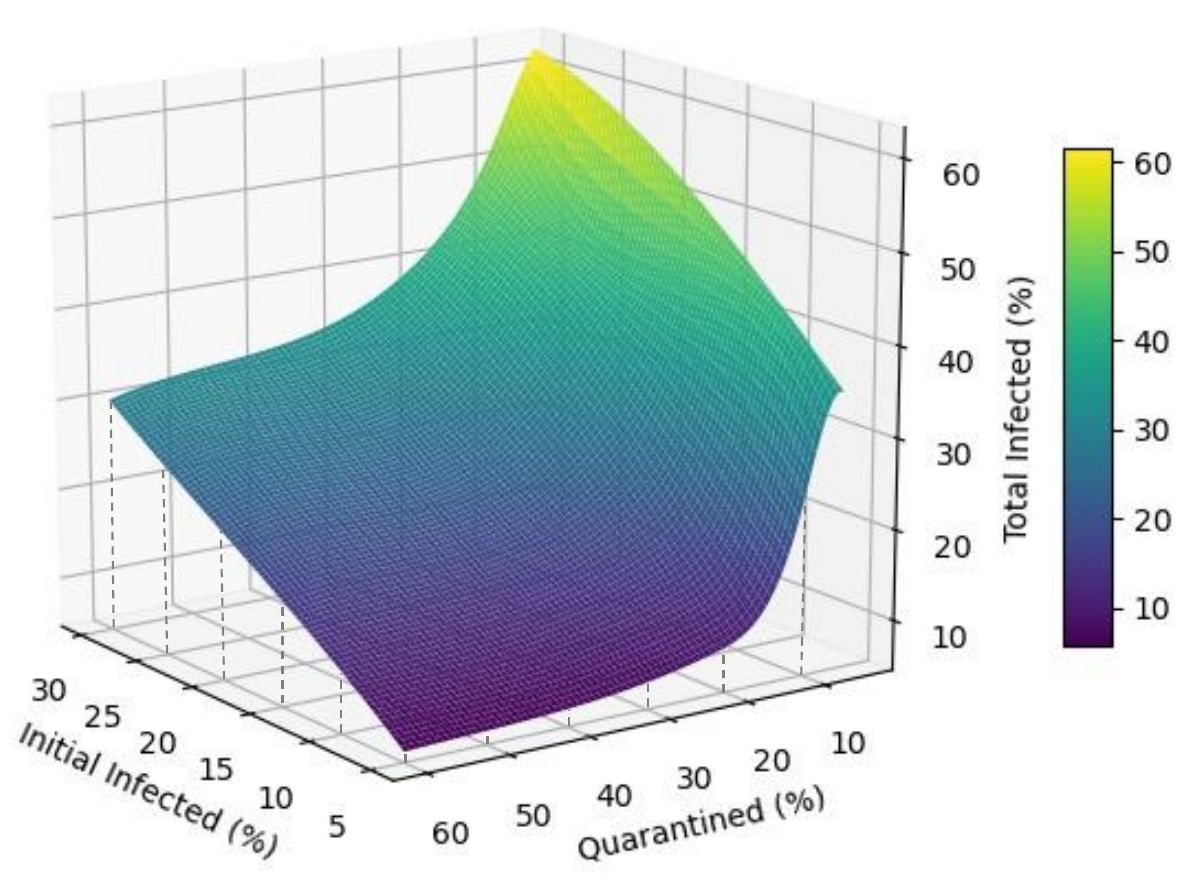}
    % \centerline{\fbox{\rule{0pt}{2in} \rule{0.9\linewidth}{0pt}}}
    \caption{\textbf{Effect of infected population in the spread of the pandemic.} We show the fraction of total infected in terms of fractions of initial infected and quarantined people.}
    \label{fig:3d}
\end{figure}

\noindent\textbf{How to perform quarantine with constrained resources?}
Next, we study suitable proportions of the population for intervention. We show the total infected fraction with respect to different amounts of infectious and quarantined populations. \autoref{fig:3d} shows the results based on the betweenness centrality measure. We observe a clear turning point where quarantining 20\% of the whole population reduces the spread of disease among all infected ratios.
This suggests that increasing the quarantine percentage over 20\% provides only marginal benefits.

\begin{figure}[t]
\begin{minipage}{.5\linewidth}
\centering
{\label{main:a}\includegraphics[scale=.25]{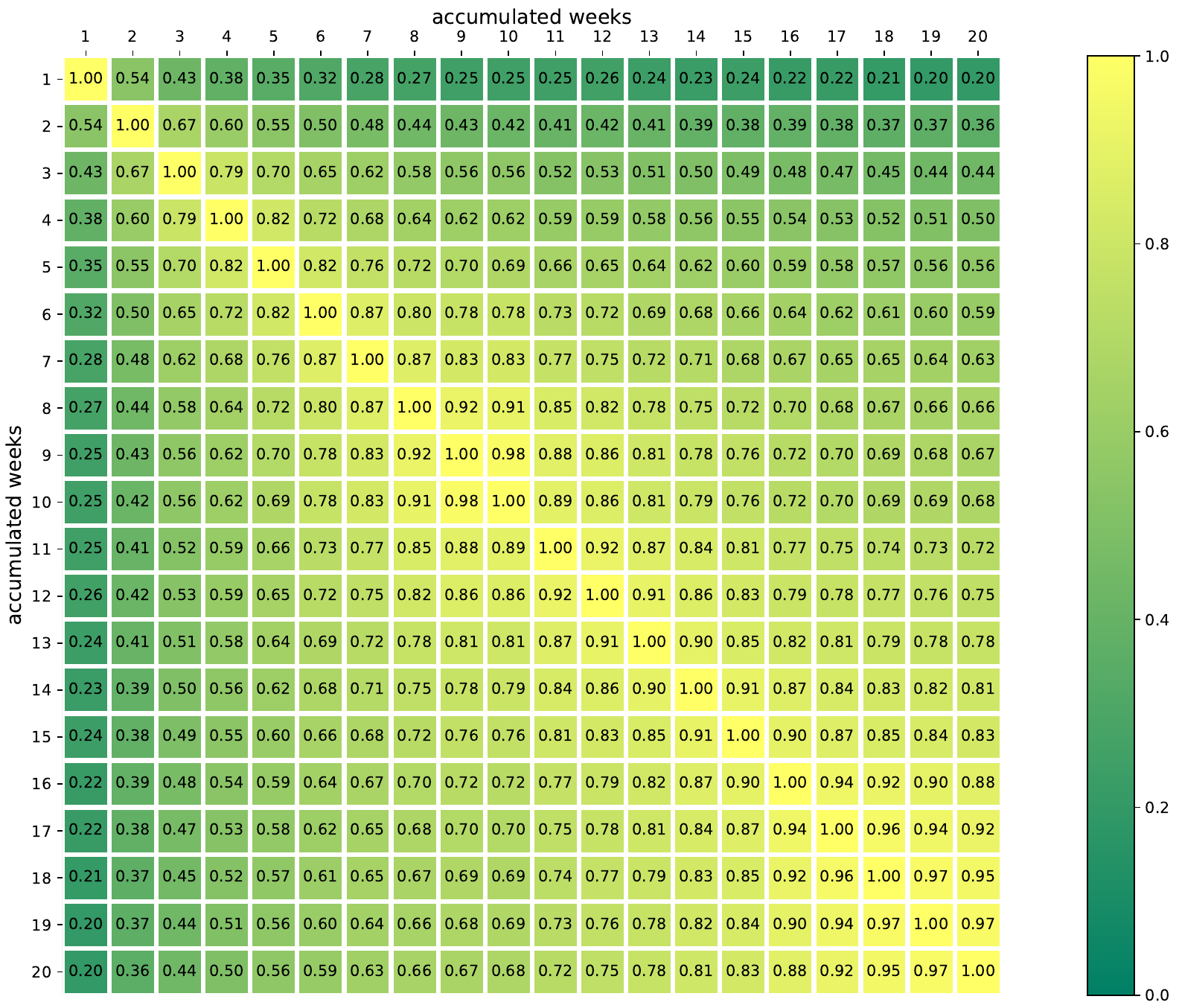}}\\
\subfloat{(a) Degree centrality}
\end{minipage}%
\begin{minipage}{.5\linewidth}
\centering
{\label{main:b}\includegraphics[scale=.25]{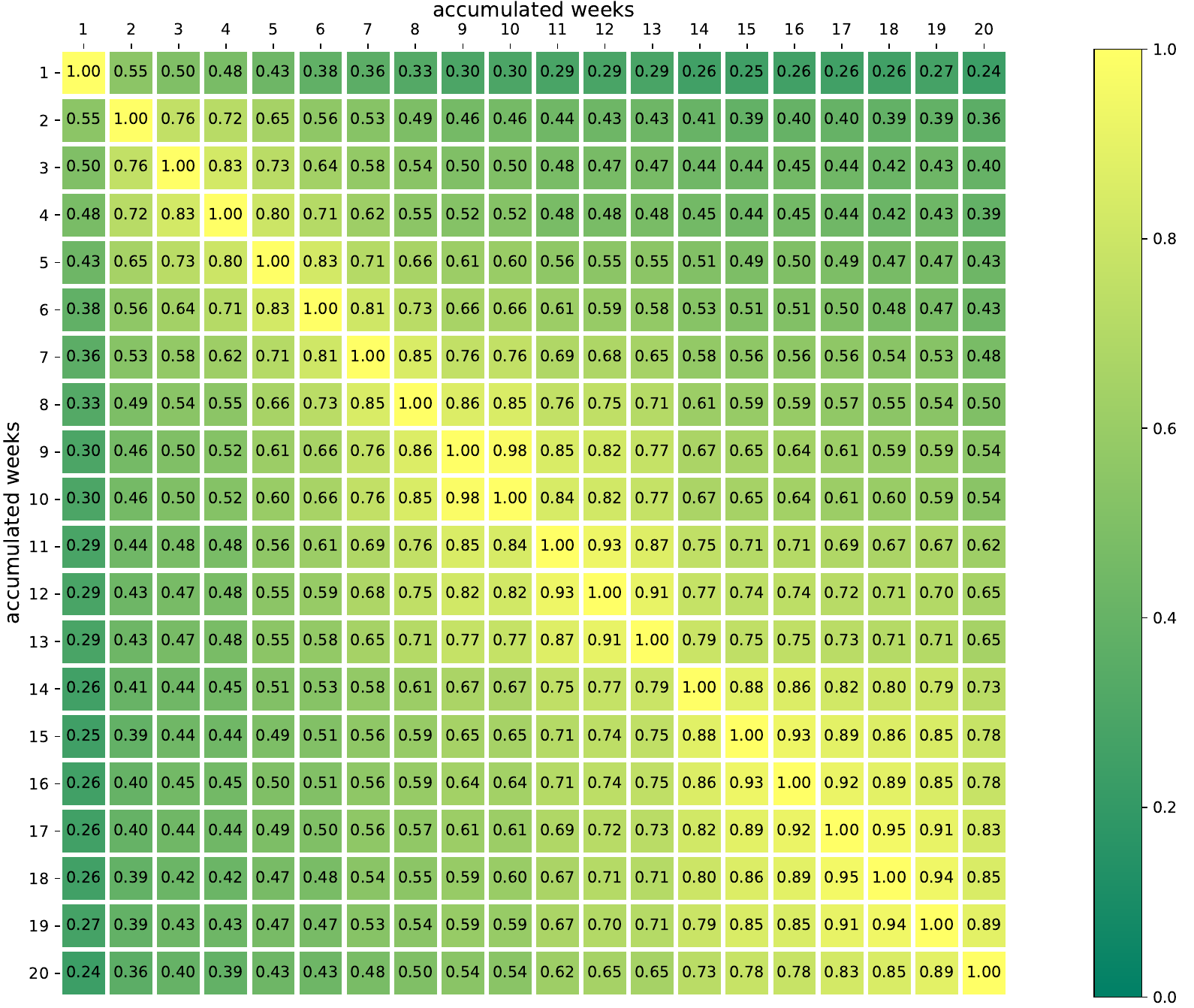}}\\
\subfloat{(b) Closeness centrality}
\end{minipage}
\par\medskip
\centering
{\label{main:c}\includegraphics[scale=.25]{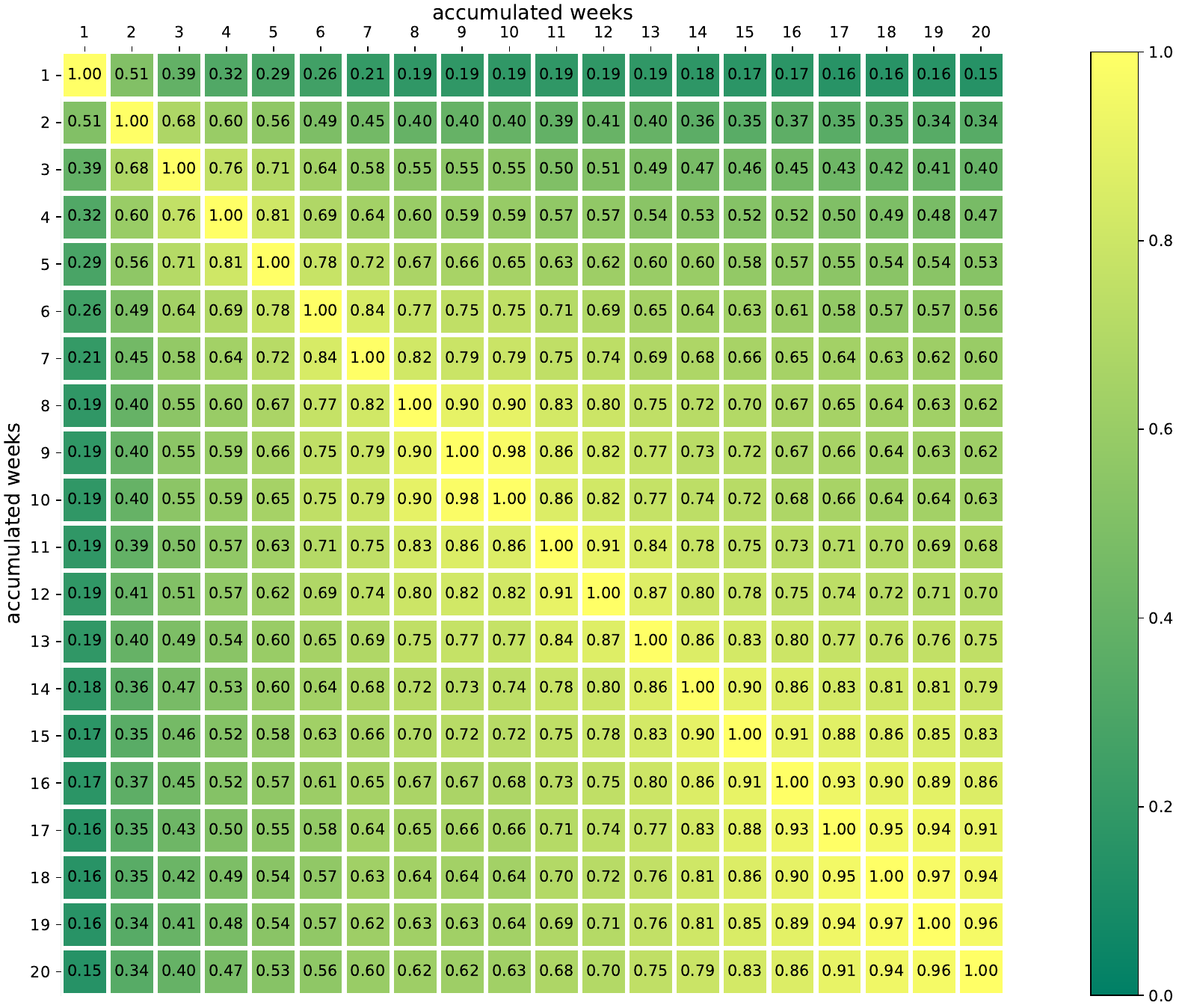}}\\
\subfloat{(c) Betweenness centrality}
\caption{\textbf{Similarity matrix of superspreaders between accumulated weeks.} }
\label{fig:accumulated_week}
\end{figure}

\noindent\textbf{Will superspreaders change during the whole semester?}
To further analyze the stability of superspreaders among different periods, we compute the similarities of the identified superspreaders from any two accumulated weeks, whose results are shown in \autoref{fig:accumulated_week}. In this study, we first generate the contact graphs based on the first N weeks in the WLAN dataset, where N ranges from 1 to 20. Next, we select the top 100 students based on our centrality measurements. We adopt rank-biased overlap (RBO)~\cite{webber2010similarity} to compute the similarity of two ranked student lists from any two accumulated weeks. Our results show that the superspreaders change during the first few weeks, but remain stable throughout the rest of the semester. For example, the similarity between the first 20 weeks and 15 weeks is around 0.8, opening up opportunities to discover the superspreaders as early as possible for efficient pandemic mitigation.

%% file: paper/5.related.tex
\subsection{Client-based Contact Tracing}
Researchers have devoted considerable attention to mobile application (app) technology for COVID-19 contact tracing. For example, Covid Watch~\cite{covidwatch} uses Bluetooth signals to detect when users are near each other and alerts them anonymously if they were in contact with someone who is later diagnosed with COVID-19. Similarly, PACT~\cite{pact} uses inter-phone Bluetooth communications (including energy measurements) as a proxy for inter-person distance measurement. Through applied cryptography, this system can collect and maintain weeks of contact events. Later, PACT augments these events with infection notifications leading to exposure notifications to all mobile phone owners who have had medically significant contact (in terms of distance and time) with infected people in the past medically significant period (e.g., two weeks). In addition, Singapore launched the TraceTogether~\cite{tracetogether} app to boost COVID-19 contact tracing efforts. By downloading the app and consenting to participate in it, TraceTogether lets users ``proactively help'' in the contact tracing process~\cite{tracetogether}. The app works by exchanging short-range Bluetooth signals between phones to detect other app users who are nearby. Apple and Google~\cite{googleandapple} are working together for the first time on a protocol that will alert users if they have been exposed to the coronavirus. Luo et al. propose A-Turf~\cite{luo2020acoustic}, an acoustic encounter detection method for COVID-19 contact tracing. Compared with Bluetooth technology, the system more precisely detects encounters within 6-foot ranges (social distancing). Unlike the WLAN-log-based contact tracing presented in this paper, client-based contact tracing requires users’ widespread adoption and active participation.

\subsection{Infrastructure-based Contact Tracing}
Infrastructure-based methods take advantage of existing infrastructure deployed worldwide such as CCTV footage~\cite{skoll2020covid}, locations measured using cellular networks~\cite{ayan2020characterizing}, Wi-Fi hotspots~\cite{trivedi2020wifitrace,zakaria2020analyzing}, and GPS~\cite{bay2020bluetrace}, without requiring client-side involvement. Similar to our approach, recent  efforts~\cite{trivedi2020wifitrace,zakaria2020analyzing} use passive Wi-Fi sensing for network-based contact tracing for infectious diseases, particularly focused on the COVID-19 pandemic. Those works mainly use location occupancy or number of contact as the measure to identify the superspreaders while we consider different types of centrality for measuring the ``importance'' of the vertex in the social networks. Moreover, we adopt SEIR simulation to justify which measure is better in discovering the superspearders.

%% file: paper/6.disc.tex
In this paper, we focused on WLAN-log-based superspreader detection in the COVID-19 pandemic. We proposed a general framework with applications to a wide range of working scenarios based on users' preferences, environmental dynamics, and resource availability. Moreover, we presented asymmetric contact, a new type of human contact. The concept of asymmetric contact partially captured the notion of environmental infection. We required that persons in asymmetric contact must have had a certain overlap time between their association times with a specific AP. In fact, we can generalize by eliminating this constraint. We can treat the overlap time as a control knob to adjust the degree of ``asymmetry''. Due to space limitations, this remains part of our future work.  We have implemented our framework, conducted an extensive evaluation, and obtained a set of important findings. Our proposed contact tracing framework and our findings provided a tool as well as guidelines for public health administrators regarding both proactive and reactive interventions against the pandemic. 
 
\section*{Acknowledgement}
The work was supported in part by the National Science Foundation (NSF) under Grant No. CNS 2028547. Any opinions, findings, conclusions, and recommendations in this paper are those of the authors and do not necessarily reflect the views of the funding agencies.